\documentclass[twocolumn,prl,amsmath,amssymb]{revtex4}

\usepackage{graphicx}
\usepackage{dcolumn}
\usepackage{bm}
\usepackage{color, float}

\begin{document}

\title{An Entanglement-Enhanced Microscope}

\author{Takafumi Ono$^{1,2}$}
\author{Ryo Okamoto$^{1,2}$}%
\author{Shigeki Takeuchi$^{1,2}$}
\email{takeuchi@es.hokudai.ac.jp}
\affiliation{%
$^1$Research Institute for Electronic Science, Hokkaido University, Sapporo 060-0812, Japan
}%
\affiliation{$^2$The Institute of Scientific and Industrial Research, Osaka University, 8-1 Mihogaoka, Ibaraki, Osaka 567-0047, Japan}


\begin{abstract}
Among the applications of optical phase measurement, the differential interference contrast microscope is widely used for the evaluation of opaque materials or biological tissues. However, the signal to noise ratio for a given light intensity is limited by the standard quantum limit (SQL), which is critical for the measurements where the probe light intensity is limited to avoid damaging the sample. The SQL can only be beaten by using {\it N} quantum correlated particles, with an improvement factor of $\sqrt{N}$.
Here we report the first demonstration of an entanglement-enhanced microscope, which is a confocal-type differential interference contrast microscope where an entangled photon pair ({\it N}=2) source is used for illumination. 
An image of a Q shape carved in relief on the glass surface is obtained with better visibility than with a classical light source. The signal to noise ratio is 1.35$\pm$0.12 times better than that limited by the SQL.
\end{abstract}

\maketitle

Quantum metrology involves using quantum mechanics to realize more precise measurements than can be achieved classically \cite{gi-sci-306-1330}. The canonical example uses entanglement of $N$ particles to measure a phase with a precision $\Delta\phi=1/N$, known as the Heisenberg limit. Such a measurement outperforms the $\Delta\phi=1/\sqrt{N}$ precision limit possible with $N$ unentangled particles---the standard quantum limit (SQL). Progress has been made with trapped ions \cite{me-prl-86-5870,le-nat-438-639,ro-nat-443-316} and atoms \cite{wi-prl-92-160406}, while high-precision optical phase measurements have many important applications, including microscopy, gravity wave detection, measurements of material properties, and medical and biological sensing. Recently, the SQL has been beaten with two photons \cite{ra-prl-65-1348,ku-qso-10-493,fo-prl-82-2868,ed-prl-89-213601,ei-prl-94-090502} and four photons \cite{Nag07,Oka08,Xia10}. 

Perhaps the natural next step is to demonstrate entanglement-enhanced metrology\cite{cr-apl-100-233704,Wol13,Mic13}. Among the applications of optical phase measurement, microscopy is essential in broad areas of science from physics to biology. The differential interference contrast microscope\cite{Nom55} (DIM) is widely used for the evaluation of opaque materials or the label-free sensing of biological tissues\cite{Jes11}. For instance, the growth of ice crystals has recently been observed with a single molecular step resolution using a laser confocal microscope combined with a DIM\cite{Saz10}. The depth resolution of such measurements is determined by the signal to noise ratio (SNR) of the measurement, and the SNR is in principle limited by the SQL. In the advanced measurements using DIM, the intensity of the probe light, focused onto a tiny area of $\sim10^{-13}$ m$^2$, is tightly limited for a noninvasive measurement, and the limit of the SNR is becoming a critical issue.

In this work, we demonstrated an entanglement-enhanced microscope, consisting of a confocal-type differential interference contrast microscope equipped with an entangled photon source as a probe light source, with an SNR of 1.35 times better than that of the SQL. We use an entangled two-photon source with a high fidelity of 98\%, resulting in a high two-photon interference visibility in the confocal microscope setup of 95.2\%. An image of a glass plate sample, where a Q shape is carved in relief on the surface with a ultra-thin step of $\sim$ 17 nm, is obtained with better visibility than with a classical light source. The improvement of the SNR is 1.35 $\pm 0.12$, which is consistent with the theoretical prediction of 1.35. We also confirm that the bias phase dependence of the SNR completely agrees with the theory without any fitting parameter. We believe this experimental demonstration is an important step towards entanglement-enhanced microscopy with ultimate sensitivity.

\begin{figure}[t] 
\begin{picture}(100,330) 
{\makebox(100,320) 
{
\scalebox{0.25}[0.25]{\includegraphics[angle=0]{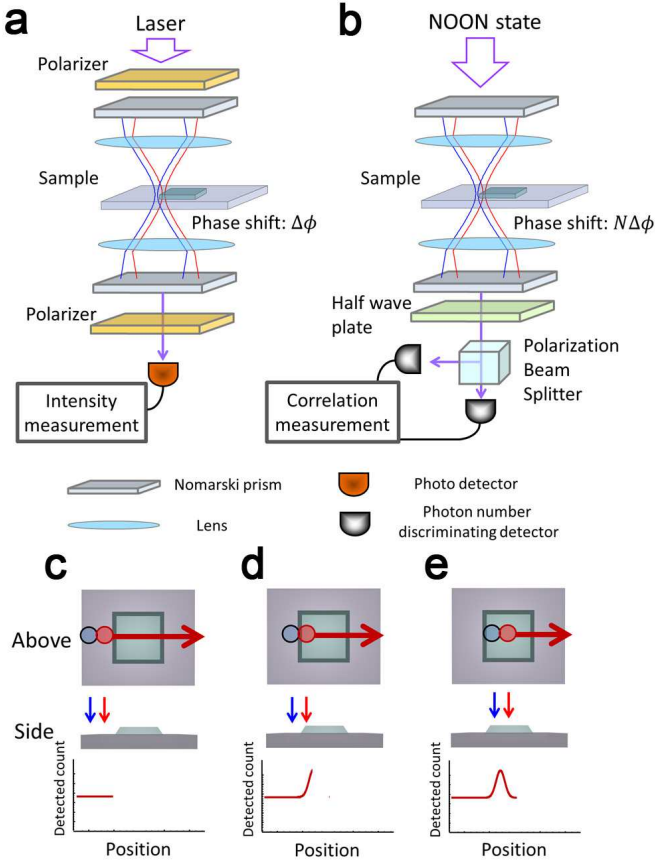}}}}
\end{picture}
\caption{\label{fig1}
Illustration of {\bf (a)} LCM-DIM and {\bf (b)} the entanglement-enhanced microscope. The red and blue lines indicate horizontally and vertically polarized light. {\bf (c), (d)} and {\bf (e)} The change in the signal while the sample is scanned.
}
\end{figure}



Our entanglement-enhanced microscope is based on a laser confocal microscope combined with a differential interference contrast microscope (LCM-DIM)\cite{Saz10,Saz12}. An LCM-DIM can detect a very tiny difference between optical path lengths in a sample. The LCM-DIM works on the principle of a polarization interferometer (Fig. 1a). In this example, the horizontal (H) and vertical (V) polarization components are directed to different optical paths by a Nomarski prism. At the sample, the two beams experience different phase shifts ($\Delta \phi_{\mathrm{H}}$ and $\Delta \phi_{\mathrm{V}}$) depending on the local refractive index and the structure of the sample. After passing through the sample, the two beams are combined into one beam by another Nomarski prism. The difference in the phase shifts can be detected as a polarization rotation at the output, $\Delta \phi = \Delta \phi_{\mathrm{V}} - \Delta \phi_{\mathrm{H}}$. 

We obtain differential interference contrast images for a sample by scanning the relative position of the focused beams on the sample (Fig. 1c to 1e). When two beams probe a homogeneous region, the output intensity is constant (Fig. 1c). At the boundary of the two regions, the signal intensity increases or decreases, since the difference in the phase shift $\Delta \phi$ becomes non-zero (Fig. 1d). The signal intensity returns to the original level after the boundary (Fig. 1e). The smallest detectable change in the phase shift is limited by the SNR, which is the ratio of the change in the signal intensity, $C(\phi)$, and the fluctuation of the uniform background level, $\Delta C$, at a bias level of $\Phi_0$. As is discussed in detail later, it is known that the SNR is limited by so-called `shot noise' or the SQL, when `classical' light sources such as lasers or lamps are used. That is, for a limited number of input photons ($N$), the SNR is limited by $\sqrt{N}$. This SNR limits the height resolution of the LCM-DIM when used to observe elementary steps at the surface of ice crystals\cite{Saz10} or the difference in refractive indexes inside a sample. Thus, improving the SNR beyond the SQL is a revolutionary advance in microscopy.

We propose to use multi-photon quantum interference to beat this standard quantum limit (Fig. 1b). Instead of a classical light, we use an entangled photon state ($(|N;0 \rangle_{\mathrm{HV}} + |0;N \rangle_{\mathrm{HV}})/\sqrt{2}$), so-called `NOON' state, which is a quantum superposition of the states `$N$ photons in the H polarization mode' and `$N$ photons in the V polarization mode'. The phase difference between these two states is $N \Delta \phi$ after passing through the sample, which is $N$ times larger than the classical case ($N=1$). At the output, the result of the multi-photon interference (the parity of the photon number in the output) is measured by a pair of photon number discriminating detectors (PNDs) \cite{Ger00,Kim12}.

As is well known from two-path interferometry with $N$ photon states, entanglement can increase the sensitivity of a phase measurement by a factor of $\sqrt{N}$. In the entanglement-enhanced microscope, we apply this effect to achieve an SNR that is $\sqrt{N}$ higher than that of the LCM-DIM. If the average number of $N$-photon states that pass through the microscope during a given time interval is $k$, then the average number of detection events in the output is given by $C(\phi) =k P(\phi)$, where $P(\phi) = ( 1 - V_N \cos(N\phi + N\Phi_0))/2$ is the probability of detecting an odd (or even) number of photons in a specific output polarization (see methods section). For small phase shifts of $\Delta \phi \ll 1/N$, the phase dependence of the signal is given by the slope of $C(\phi)$ at the bias phase $\Phi_0$, $\mathrm{lim}_{\Delta \phi \to 0}  C(\Delta \phi) - C(0) = \left|\partial C(\phi)/\partial \phi\right|_{\phi=0} \times \Delta \phi$, and the SNR is given by the ratio of the slope and the statistical noise of the detection. If the emission of the $N$-photon states is statistically independent, the statistical noise is given by $\Delta C|_{\phi=0} = \sqrt{kP(0)}$. The SNR is then given by
\begin{equation}
\label{SNR}
\mathrm{SNR} =(1-\xi) \sqrt{\frac{k}{2}}N V_N \frac{\left| \sin \left( N\Phi_0 \right) \right|}{\sqrt{1-V_N \cos \left( N\Phi_0 \right)}} \times \Delta \phi,
\end{equation}
where $\xi$ is the normalized overlap region between the two beams at the sample plane (see methods section). 
By maximizing the function of $\cos \left( N\Psi_0 \right)$, we can find the maximum sensitivity ($\mathrm{SNR}_{\max}$) as follows:
\begin{equation}
\label{SNRmax}
\mathrm{SNR}_{\max} = (1-\xi) \sqrt{k}N \sqrt{1-\sqrt{1-V_N^2}} \times \Delta \phi,
\end{equation}
at a bias phase of 
\begin{equation}
\label{bias}
\cos (N \Phi_0)=\left( 1-\sqrt{1-V_N^2}\right)/V_N.
\end{equation}
For the ideal case of $V_N=1$, the maximum sensitivity $\mathrm{SNR}_{\max}=\sqrt{k} N \times \Delta \phi$. Since the SNR for a classical microscope using $kN$ photons $\mathrm{SNR}_{\max} = \sqrt{kN} \times \Delta \phi$, an entanglement-enhanced microscope can improve the SNR by a factor of $\sqrt{N}$ compared to a classical microscope for the same photon number.

\begin{figure}[t] 
\begin{picture}(200,140) 
{\makebox(200,60) 
{
\scalebox{0.1}[0.1]{\includegraphics[angle=0]{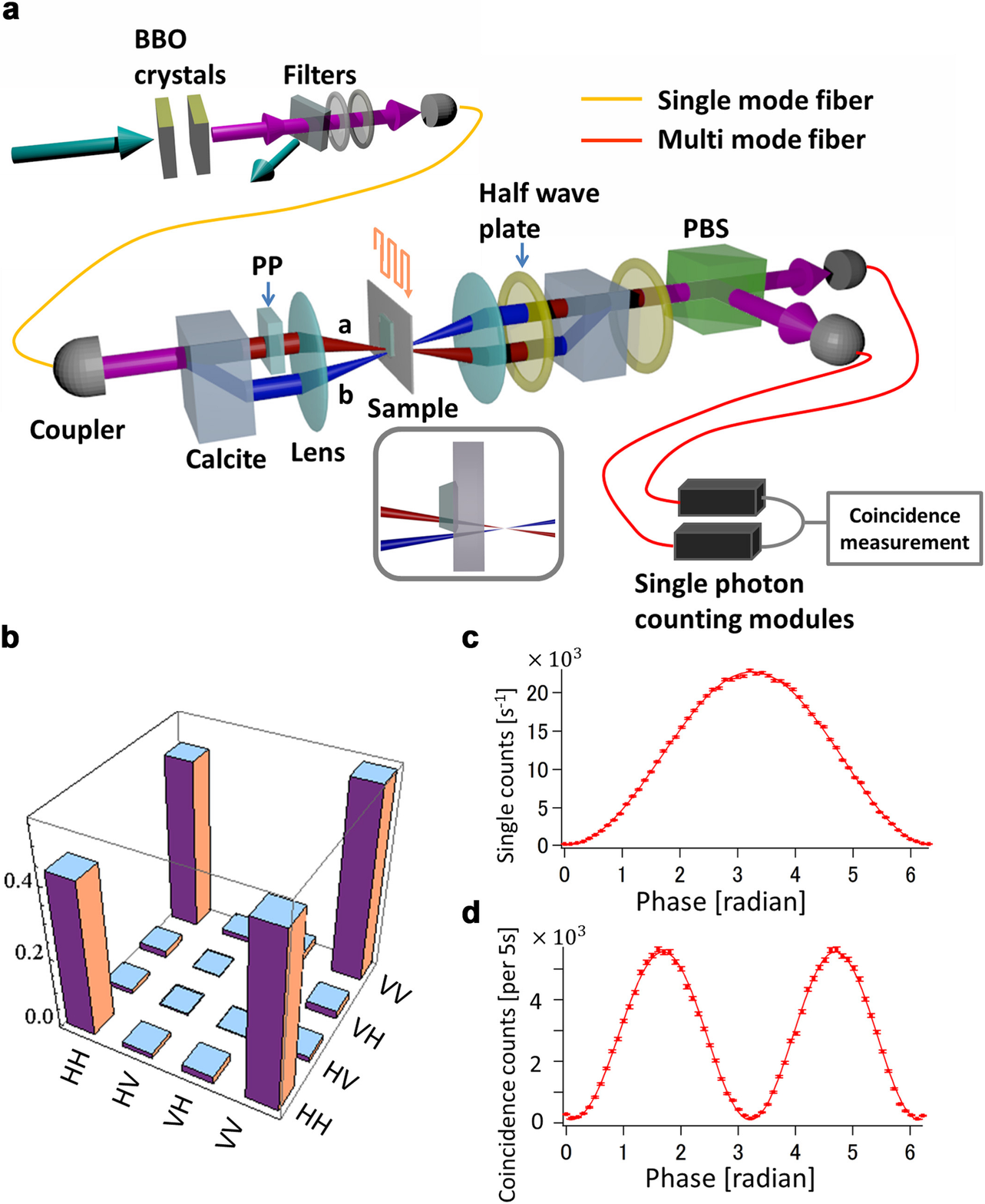}}}}
\end{picture}
\vspace{4cm}
\caption{\label{fig2}
({\bf a}) Experimental setup for a two-photon entanglement-enhanced microscope. A 405-nm diode laser (line width $<$ 0.02 nm) was used for the pump beam. A sharp cut filter with a cutoff wavelength below 715 nm and a band pass filter with 4-nm bandwidth were used. The beam displacement at the calcite crystal was 4 mm. The blue and red lines indicate the optical paths for the horizontally and vertically polarized beams. The inset shows an illustration of the sample with the trajectory of the two beams. ({\bf b}) Single photon interference fringes of a classical microscope [cps] after the dark counts of the detectors (100 cps) subtracted. ({\bf c}) Two-photon interference fringes of an entanglement-enhanced microscope [counts per 5 s]. While varying the phase by rotating the phase plate (PP), we counted the detected events in the output. The classical fringe was measured by inserting a polarizer transmitting the diagonally polarized photons and counting single-photon detection events.
}
\end{figure}

We demonstrate an entanglement-enhanced microscope (Fig. 2a) using a two-photon NOON state ($N=2$). First, a polarization entangled state of photons $\left( |2 ; 0 \rangle_{\mathrm{HV}} + |0 ; 2 \rangle_{\mathrm{HV}} \right)/\sqrt{2}$ is generated from two beta barium borate (BBO) crystals \cite{Kwi99,Whi99} and is then delivered to the microscope setup via a single-mode fiber. The polarization entangled state is then converted to a two-photon NOON state $\left( |2 \rangle_a | 0 \rangle_b + |0 \rangle_a |2 \rangle_b \right)/\sqrt{2}$ using a calcite crystal\cite{Kaw07} and focused by an objective lens. From the result of quantum state tomography (Fig. 2b), the fidelity of the state is 98 \% and the entanglement concurrence is  0.979, which ensures that the produced state is almost maximally entangled. The entangled photons pass through two neighboring spots at the sample plane (Fig. 2a inset). Then, after passing through the collimating lens, the two paths are merged by a polarization beam splitter and the result of the two photon interference is detected by a pair of single-photon counters and a coincidence counter. The sample is scanned by a motorized stage to obtain an image. The beam diameters at the sample plane and the distance between the center of the beams are all 45 $\mu$m ($\xi = 0.046$).

Figures 2c and 2d show the single-photon and two-photon interference fringes using a classical light source and the NOON source respectively. The fringe period of Fig. 2d is half that of Fig. 2c, which is a typical feature of NOON state interference. The visibilities of these fringes, $V_\mathrm{c}=97.1 \pm 0.4 \%$ (Fig. 2c) and $V_\mathrm{q}=95.2 \pm 0.6 \%$ (Fig. 2d), suggests the high quality of the classical and quantum interferences.

\begin{figure}[t]
\begin{picture}(160,220) 
{\makebox(160,180) 
{
\scalebox{0.1}[0.1]{\includegraphics[angle=0]{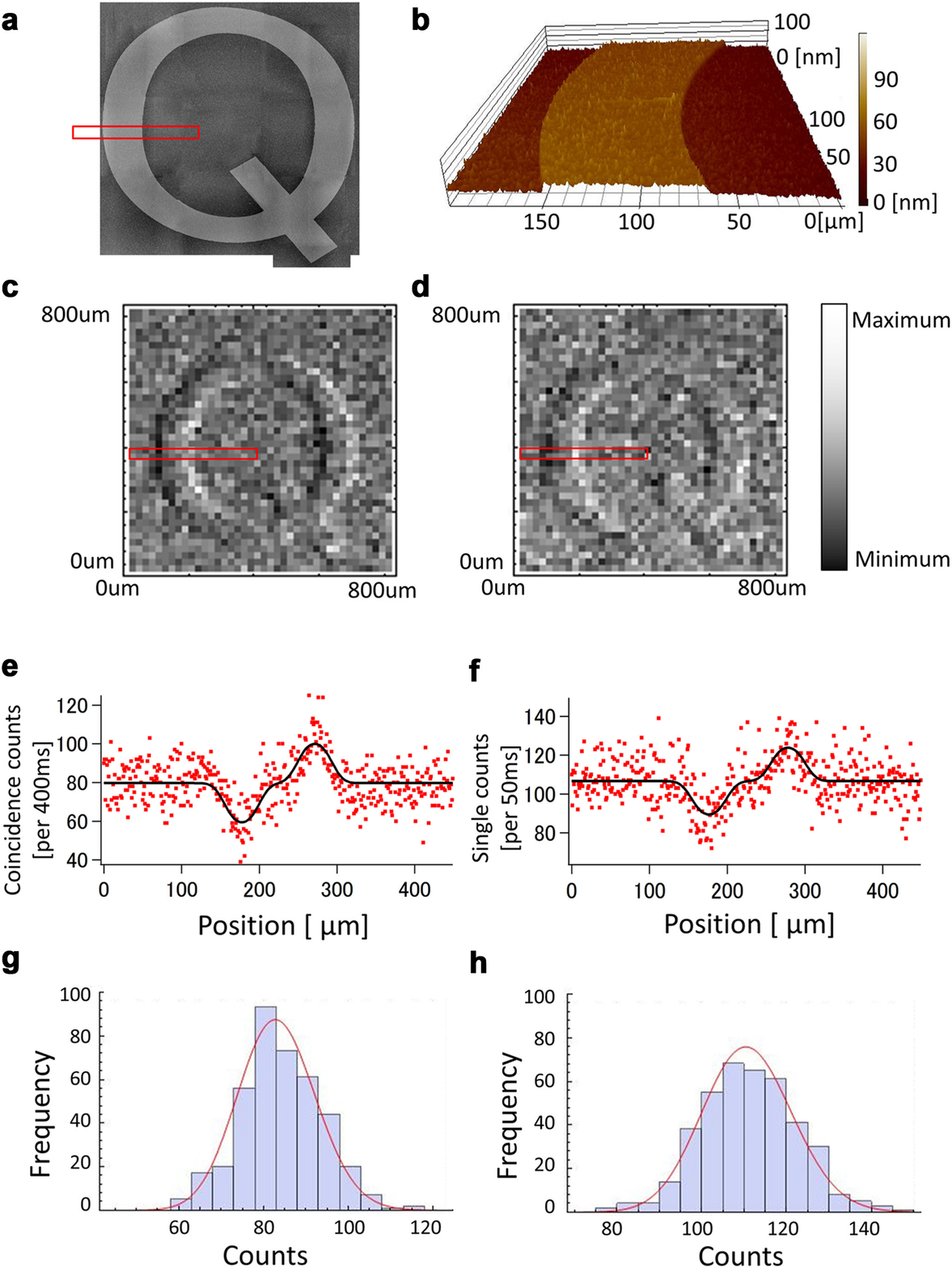}}}}
\end{picture}
\vspace{1.5cm}
\caption{\label{fig3}
({\bf a}) Atomic force microscope (AFM) image of a glass plate sample (BK7) on whose surface a `Q' shape is carved in relief with an ultra-thin step using optical lithography. ({\bf b}) The section of the AFM image of the sample, which is the area outlined in red in ({\bf a}). The height of the step is estimated to be 17.3 nm from this data. ({\bf c}) The image of the sample using an entanglement-enhanced microscope where two photon entangled state is used to illuminate the sample. ({\bf d}) The image of the sample using single photons (a classical light source). ({\bf e}) and ({\bf f}) are 1D fine scan data for the area outlined in red in ({\bf c}) and ({\bf d}) for the same photon number of 920. The measurement was made at a bias phase of 0.41 ({\bf e}) and 0.66 ({\bf f}), where optimal bias phase are 0.38 and 0.67 respectively.
}
\end{figure}

Figure 3 shows the main result of this experiment. We used a glass plate sample (BK7) on whose surface a Q shape is carved in relief with an ultra-thin step of approximately 17 nm using optical lithography (Fig. 3). Figures 3c and 3d show the 2D scan images of the sample using entangled photons and single photons, respectively. The step of the Q-shaped relief is clearly seen in Fig. 3c, while it is obscure in Fig. 3d. Note that for both images we set the bias phases to almost their optimum values given by Eq. (\ref{bias}) and the average total number of photons ($N \times k$) contributed to these data are set to 920 per position assuming the unity detection efficiency.

For more detailed analysis, the cross section of the images (coincidence count rate/single count rate at each position) are shown in Figs. 3e and 3f. The solid lines are theoretical fits to the data where the height and position of the step and the background level are used as free parameters. For Fig. 3e, the signal (the height of the peak of the fitting curve from the background level) is $20.21 \pm 1.13$, and the noise ( the standard deviation of each experimental counts from the background level of the fitting curve ) is $9.48$ (Fig. 3g). Thus, the SNR is $2.13 \pm 0.12$. Similarly, the signal, the noise, and the SNR are $17.7 \pm 1.22$, $11.25$ (Fig. 3h), and $1.58 \pm 0.11$ for Fig. 3f where classical light source (single photons) are used. The improvement in SNR is thus $1.35 \pm 0.12$, which is consistent with the theoretical prediction of $1.35$ (Eq. 2). The estimated height of the step was $17.0 \pm 0.9$ nm (quantum) and $16.6 \pm 1.1$ nm (classical), and is consistent with the estimated value of 17.3 nm from AFM image in Fig. 3b.

As Eq. (\ref{SNR}) shows, the SNR depends on the bias phase. Finally, we test the theoretical prediction of the bias phase dependence given by Eq. (\ref{SNR}) in actual experiments. Figure 4 shows the bias phase dependence of the SNR for the two-photon NOON source (Fig. 4a) and classical light source (Fig. 4b). The solid curve is the theoretical prediction calculated by Eq. (\ref{SNR}), where we used the observed visibilities of the fringes in Figs. 2c and 2d for $V_N$. The theoretical curves are in good agreement with the experimental results.

Note that the entanglement-enhanced microscope we reported here is different from the `entangled-photon microscopef, which is the combination of two-photon fluorescence microscopy and the entangled photon source, theoretically proposed by Teich and Saleh\cite{Tei97}. In the proposal, the increase in two photon absorption rate and the flexibility in the selection of target regions in the specimen were predicted. The application of entangled photon sources for imaging also includes quantum lithography\cite{Bot01}, where the lateral resolution of the generated pattern is improved\cite{Kaw07,DAn01}, and ghost imaging\cite{Pit95}, where the spatial correlation of entangled photons is utilized. In this context, this work is the first application of entanglement-enhanced optical phase measurement beyond the SQL for imaging including microscopy. Note also that the entanglement is indispensable to improve the SNR of the phase measurement beyond the SQL \cite{Gio06,Nag07}. This situation is different from the improvement in the contrast of the ghost imaging using strong thermal light\cite{Ou07,CaoD08,Liu09,Cha09,Cha10,Che10,CaoB10}.

\begin{figure}[t]
\begin{picture}(180,100) 
{\makebox(190,100) 
{
\scalebox{0.25}[0.25]{\includegraphics[angle=0]{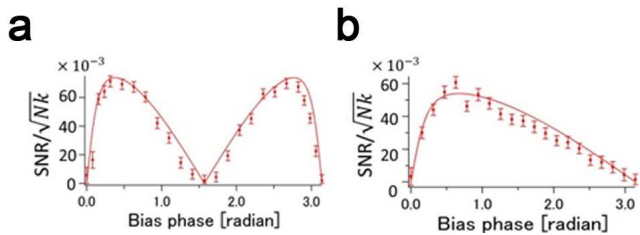}}}}
\end{picture}
\vspace{-0.1cm}
\caption{\label{fig4}
Dependence of the SNR on the bias phase using ({\bf a}) an entanglement-enhanced microscope ($N=2$) and ({\bf b}) a classical microscope ($N=1$). The SNR was calculated from the experimental data similar to Fig. 3E and 3F taken for different bias phases. The total photon number $Nk=1150$ for ({\bf a}) and $1299$ for ({\bf b}). The solid line shows the theoretical curve using Eq. (1).}
\end{figure}

In conclusion, we proposed and demonstrated an entanglement-enhanced microscope, which is a confocal-type differential interference contrast microscope equipped with an entangled photon source as a probe light source, with an SNR 1.35 $\pm 0.12$ times better than the SQL. Imaging of a glass plate sample with an ultra-thin step of $\sim$ 17 nm under a low photon number condition shows the viability of the entanglement-enhanced microscope for light sensitive samples. To test the performance of the entanglement-enhanced microscope, we used modest-efficiency detectors, however, recently developed high-efficiency number-resolving photon detectors would markedly improve detection efficiency\cite{Tak99,Fuk11}. We believe this experimental demonstration is an important step towards entanglement-enhanced microscopy with ultimate sensitivity, using a higher NOON state, a squeezed state\cite{go-natphys-4-472,Ani10}, and other hybrid approaches\cite{Ono10} or adaptive estimation schemes\cite{Oka12,Han02}.\\

\section*{Acknowledgment}
The authors thank Prof. Mitsuo Takeda, Prof. Yoko Miyamoto, Prof. Hidekazu Tanaka, Prof. Masamichi Yamanishi, Dr. Shouichi Sakakihara, and Dr. Kouichi Okada. This work was supported in part by FIRST of JSPS, Quantum Cybernetics of JSPS, a Grant-in-Aid from JSPS, JST-CREST, Special Coordination Funds for Promoting Science and Technology, Research Foundation for Opto-Science and Technology, and the GCOE program.


\newpage
\appendix

\section{APPENDIX}
To perform the parity measurement used in (ref. 21,22), it is required to count both of the events where `even' and `odd' number of photons are detected in the output. However, experimental implementations become much easier if it is sufficient for us to just count `odd' (or `even') number photon-detection events. In addition, the distance between the two beams and the beam size may effect on the SNR. Here we consider these technical effects on SNR and derive Eq.(1).

To derive Eq. (1), we consider that the two beams at the sample are separated at a distance of $\alpha$ along the $x$-axis, and each set of $N$ photons has Gaussian distribution with a variance of $\sigma$ in the $x$-$y$ plane (Fig. 5).
After passing through the sample, the two beams experience a phase shift ($\phi$) in the gray region ($x>0$). The state, $|\Psi(x,y,\phi) \rangle$, after the sample is written as
\begin{equation}
|\Psi(x,y,\phi) \rangle = \frac{1}{\sqrt{2}} \left( |\psi_{\mathrm{H}}(x,y,\phi) \rangle + \mathrm{e}^{iN\Phi_0}|\psi_{\mathrm{V}} (x,y,\phi) \rangle \right),
\end{equation}
where $|\psi_{\mathrm{H}}(x,y,\phi) \rangle$ and $|\psi_{\mathrm{V}}(x,y,\phi) \rangle$ represents the states of $N$ photons in the horizontal and vertical polarization modes respectively, and $\Phi_0$ is a bias phase. We assume that the phase shift is described by a step function and the $N$ photons are in the same spatial modes. These states are written as 

\begin{widetext}
\begin{eqnarray}
|\psi_{\mathrm{H}} (x,y,\phi) \rangle &=& \int^{\infty}_{-\infty} \int^{\infty}_{-\infty}~ \mathrm{e}^{-i\chi(x) N\phi}\sqrt{f(x-\alpha/2,y)} \frac{1}{\sqrt{N!}} \left( \hat{a}^{\dagger}_{\mathrm{H}}(x,y) \right)^N |0 \rangle~ dx dy \nonumber\\ 
                             &=& \int^{\infty}_{-\infty} \int^{\infty}_{-\infty}~ \mathrm{e}^{-i\chi(x) N\phi}\sqrt{f(x-\alpha/2,y)} |N;0, x,y \rangle_{\mathrm{HV}} ~dx dy, \nonumber\\
|\psi_{\mathrm{V}} (x,y,\phi) \rangle &=& \int^{\infty}_{-\infty} \int^{\infty}_{-\infty}~ \mathrm{e}^{-i\chi(x) N\phi} \sqrt{f(x+\alpha/2,y)} \frac{1}{\sqrt{N!}} \left( \hat{a}^{\dagger}_{\mathrm{V}}(x,y) \right)^N |0 \rangle~ dx dy \nonumber\\
                             &=&\int^{\infty}_{-\infty} \int^{\infty}_{-\infty}~ \mathrm{e}^{-i\chi(x) N\phi} \sqrt{f(x+\alpha/2,y)} |0;N, x,y \rangle_{\mathrm{HV}} ~dx dy , 
\end{eqnarray}
\end{widetext}
where $|0 \rangle $ is a vacuum state, $\hat{a}^{\dagger}_\mathrm{H}(x,y)$ and $\hat{a}^{\dagger}_\mathrm{V}(x,y)$ are the creation operators in H and V polarization modes at the position of $(x,y)$ respectively, and $\chi(x)$ is a step function that $\chi(x)=0$ for $x \leq 0$ and $\chi(x)=1$ for $x > 0$. $f(x-\alpha/2,y)$ and $f(x+\alpha/2,y)$ represent the $N$ photon probability densities in the horizontal and vertical polarization modes written as
\begin{equation}
f(x,y)             = \frac{1}{2\pi \sigma^2} \mathrm{e}^{-\frac{(x^2+y^2)}{2\sigma^2}};~\int^{\infty}_{-\infty} \int^{\infty}_{-\infty}~ f(x,y) ~dx dy = 1.
\end{equation}
After passing through the second calcite crystal, the two beams are displaced a distance of $- \alpha/2$ for H polarization and $\alpha/2$ for V polarization along the $x$-axis, resulting in the two beams in the same spatial mode. The state can then be written as
\begin{widetext}
\begin{equation}
|\Psi' (x,y,\phi) \rangle = \frac{1}{\sqrt{2}} \left( |\psi_{\mathrm{H}}(x+\alpha/2,y,\phi) \rangle + \mathrm{e}^{iN\Phi_0}|\psi_{\mathrm{V}} (x-\alpha/2,y,\phi) \rangle \right).
\end{equation}
\end{widetext}
Here, we assume that the state is projected onto the state where odd number of photons are in the minus diagonal polarization mode at the output. The measurement operator in the basis of plus (P) and minus (M) diagonal polarization is therefore written as 
\begin{widetext}
\begin{eqnarray}
\hat{\Pi} =
\begin{cases}
\int^{\infty}_{-\infty} \int^{\infty}_{-\infty}~ \sum_{n=0}^{\infty}~ \sum_{m=1}^{N/2}~ |n; 2m-1,x,y \rangle_{\mathrm{PM}} ~_{\mathrm{PM}}\langle n; 2m-1,x,y| ~dx dy & \mathrm{if}~ N~ \mathrm{is~ even} \\
 & \\[-0.4cm]
\int^{\infty}_{-\infty} \int^{\infty}_{-\infty}~ \sum_{n=0}^{\infty}~ \sum_{m=1}^{(N+1)/2}~ |n; 2m-1,x,y \rangle_{\mathrm{PM}} ~_{\mathrm{PM}}\langle n; 2m-1,x,y| ~dx dy & \mathrm{if}~ N~ \mathrm{is~ odd.}
\end{cases}
\end{eqnarray}
\end{widetext}
where
\begin{widetext} 
\begin{eqnarray}
|n;m, x,y \rangle_{\mathrm{PM}} &=& \frac{1}{\sqrt{n! m!}} \left( \hat{a}^{\dagger}_{\mathrm{P}}(x,y) \right)^n \left( \hat{a}^{\dagger}_{\mathrm{M}}(x,y) \right)^m |0 \rangle \nonumber\\
                        &=& \frac{1}{\sqrt{n!m!}} \left( \frac{1}{\sqrt{2}} \left( \hat{a}^{\dagger}_{\mathrm{H}}(x,y) + \hat{a}^{\dagger}_{\mathrm{V}}(x,y) \right) \right)^n \left( \frac{1}{\sqrt{2}} \left( \hat{a}^{\dagger}_{\mathrm{H}}(x,y) - \hat{a}^{\dagger}_{\mathrm{V}}(x,y) \right) \right)^m |0 \rangle.\nonumber\\
\end{eqnarray}
\end{widetext}
The probability of odd number photon-detection can then be written
\begin{widetext}
\begin{eqnarray}
\label{probability}
P(\phi) &=& \langle \Psi'(x,y,\phi) | \hat{\Pi} | \Psi'(x,y,\phi) \rangle \nonumber\\
        &=& \frac{1}{2}(1-\cos(N\phi+N\Phi_0))(1-\xi(\alpha)) + \frac{1}{2}(1-\cos(N\Phi_0))\xi(\alpha),
\end{eqnarray}
\end{widetext}
where we denote the phase independent term of $\int^{\infty}_{\infty} \int^{-\alpha/2}_{-\infty} f(x,y) ~dxdy + \int^{\infty}_{\infty} \int^{\infty}_{\alpha/2} ~ f(x,y) ~dxdy$ as $\xi(\alpha)$ which is the overlap integral between H and V polarized beams at the sample plane. 

We now calculate the SNR of our microscope using the NOON state including the effect of the overlap between the two beams at the sample plane. If the emission of the $N$-photon states is statistically independent, the statistical noise at the bias phase $\Delta C(\phi)|_{\phi=0} = \sqrt{k P(0)}$ as given by
\begin{equation}
\Delta C(0) = \sqrt{k} \sqrt{\frac{1}{2}(1-\cos(N\Phi_0))} ~.
\end{equation}
For a small phase shift of $\Delta \phi \ll 1/N$, the signal is
\begin{equation}
\left. \left| \partial (C(\phi))/\partial \phi \right| \right|_{\phi=0} \times \Delta \phi =(1-\xi(\alpha)) \frac{k}{2} N \left| \sin (N\Phi_0) \right| \times \Delta \phi .
\end{equation}
Considering the visibility of the interference fringe $V_N$, the SNR is given by 
\begin{equation}
\label{SNR2}
\mathrm{SNR} = (1-\xi)\sqrt{\frac{k}{2}}N V_N \frac{\left| \sin \left( N\Phi_0 \right) \right|}{\sqrt{1-V_N \cos \left( N\Phi_0 \right)}} \times \Delta \phi.
\end{equation}
Thus one can confirm that counting odd (or even) number photon-detection events can also achieve the phase super sensitivity. Note also that the dependence of SNR on the size and the distance between the two beams is simply given by $(1-\xi)$. This means that it is reasonable to compare the SNR between an entanglement-enhanced microscope and a classical microscope ($N=1$) for the same $\xi$. 

\newpage
~
\vspace{2cm}
\begin{figure}[h]
\begin{picture}(180,100) 
{\makebox(190,100) 
{
\scalebox{0.25}[0.25]{\includegraphics[angle=0]{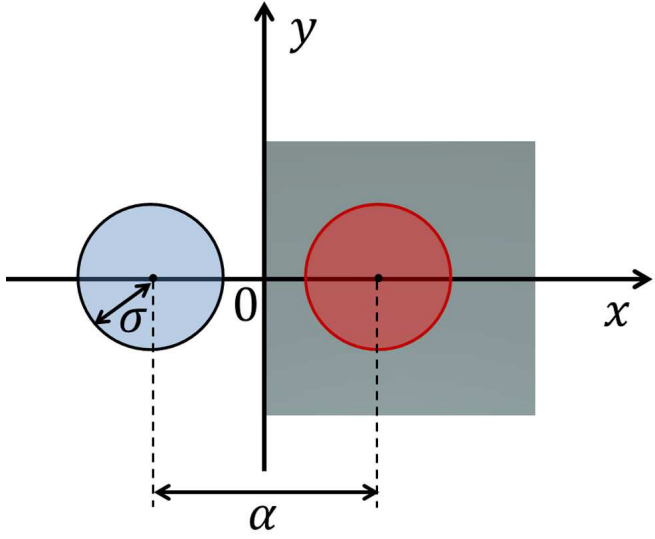}}}}
\end{picture}
\vspace{2cm}
\caption{\label{fig5}
The schematic of the two probe beams. The beams are in V polarization (blue) and H polarization (red) on the sample, corresponding to Figs 1c - 1e. The gray shaded region has the phase change of $\phi$.
}
\end{figure}

\end{document}